\documentclass{article}
\usepackage[utf8]{inputenc}
\usepackage[brazil]{babel}
\usepackage{multicol}
\usepackage{graphicx}
\usepackage{hyperref}

\title{Análise Estática de Código-Fonte}
\author{Joenio Marques da Costa\\
  joenio@joenio.me
}

\begin{document}

\maketitle

\section{Introdução}

Este artigo apresenta um resumo teórico sobre análise estática de código-fonte,
passando por definição, principais usos e aplicações, detalhes de como análise
estática é realizada, seus formatos de representação intermediário, modelos e
técnicas de análise mais comuns, finaliza apresentando um conjunto de
ferramentas de análise estática de livre acesso e disponíveis livremente para
download, ferramentas de software acadêmico desenvolvidas por cientistas
durante seus trabalhos de pesquisa e que ainda possuem baixo reconhecimento
entre seus pares \cite{costa2018sustainability}, portanto, sendo necessária sua
divulgação.

\begin{center}
\line(1,0){100}
\end{center}

\textit{Este trabalho está licenciado sob a Licença Atribuição-CompartilhaIgual 4.0
Internacional Creative Commons. Para visualizar uma cópia desta licença, visite
\url{http://creativecommons.org/licenses/by-sa/4.0/} ou mande uma carta para Creative
Commons, PO Box 1866, Mountain View, CA 94042, USA.}

\textit{O código-fonte deste artigo escrito em \LaTeX está disponível em
\url{http://gitlab.com/joenio/analise-estatica}.}

\section{Análise estática de código-fonte} \label{analise-estatica}

A análise estática de código-fonte é o primeiro passo para coletar informações
necessárias em diversas atividades de verificação, medição e melhoria da
qualidade de produtos de software \cite{Cruz2009, Kirkov2010}. Ela é
realizada com base no código-fonte de um programa ou sistema de software, e a
partir daí descobre problemas e propriedades de sua qualidade estrutural
\cite{Chess2007}.

Ferramentas de análise estática estão disponíveis há décadas, em especial,
para programadores. A ferramenta Lint \cite{Johnson1978}, considerada a
primeira ferramenta de análise estática \cite{Gosain2015}, foi criada para
examinar programas escritos em linguagem C e aplicar regras de tipagem mais
estritas do que as regras dos próprios compiladores da linguagem.

Análise estática de código-fonte tem como objetivo prover
informações acerca de um programa a partir do seu código-fonte sem
necessidade de execução, e sem requerer qualquer outro artefato do programa
além do próprio código.

É um ramo que possui muitas das suas abordagens em comum com os estudos da
área de análise de programas (\textit{program analysis}), especialmente na área de
compiladores, onde atua especialmente nas primeiras etapas do processo de compilação.

A análise estática de código-fonte é considerada uma atividade meio com
objetivo de suportar uma variedade de tarefas comuns da engenharia de
software; muitas dessas tarefas são substancialmente úteis em atividades de
manutenção. \cite{Binkley2007} define uma extensa lista dessas
atividades, incluindo:

\begin{multicols}{2}
  \begin{itemize}
    \item Análise de performance
    \item Compreensão de programas
    \item Desenvolvimento baseado em modelos
    \item Detecção de clones
    \item Evolução de software
    \item Garantia de qualidade
    \item Localizaçao de falhas
    \item Manutenção de software
    \item Recuperação arquitetural
    \item Testes
  \end{itemize}
\end{multicols}

Seja em qual atividade for, a análise estática possui uma importância
significativa, pois ao ser capaz de extrair informações diretamente do
código-fonte de um programa, pode auxiliar a responder perguntas necessárias
para as diversas atividades de desenvolvimento e evolução de software. Esta
importância se torna ainda mais aparente diante da ``lei'' da tendência para
execução \cite{Harman2010} que indica que todos os tipos de notação tem a
tendência de se tornar executáveis.

\section{Usos da análise estática de código-fonte} \label{usos}

A análise de programas trata, de modo geral, da descoberta de problemas e
fatos sobre programas, ela pode ser realizada sem a necessidade de executar o
programa (análise estática) ou com informações provenientes de sua execução
(análise dinâmica).

A idéia de que programas de computador podem ser utilizados para analisar
código-fonte de outros programas tem uma história de mais de 40 anos.  O
programa PFORT \cite{Ryder1974} foi projetado para localizar potenciais
problemas na portabilidade de código Fortran; em função da diversidade de
dialetos de Fortran, uma compilação sem erros não indicava que o programa
estava correto segundo os padrões da linguagem \cite{Wichmann1995}.

Desde então, ferramentas de análise estática de código-fonte têm surgido para
os mais diversos fins -- muitas delas a partir das pesquisas e desenvolvimentos
da área de compiladores.  O \textit{parser} utilizado nessas ferramentas têm
funcionalidades análogas aos analisadores usados em compiladores
\cite{Anderson2008}.

O uso de tais ferramentas tem se tornado mais e mais comum no ciclo de
desenvolvimento de software, sendo aplicadas em uma infinidade de atividades
distintas visto que o campo de aplicação destas ferramentas é bastante variado,
cobrindo diferentes objetivos e atividades \cite{Chess2007}.

\subsection{Verificação de tipos}

A forma mais amplamente utilizada de análise estática, e uma das quais os
programadores estão mais familiarizados, é a checagem de tipo.  Programadores
dão pouca atenção a isto, visto que as regras são definidas pela linguagem de
programação e executadas pelo próprio compilador, de forma que não se torna
necessário entender como a análise acontece.  No entanto, esta atividade de
verificação é análise estática e elimina toda uma categoria de erros de
programação. Por exemplo, previne que programadores acidentalmente atribuam
valores de forma incorreta a variáveis.  Ainda, ao capturar erros em tempo de
compilação, esta checagem de tipo previne erros em tempo de execução.

\subsection{Verificação de estilo}

Os verificadores de estilo são um tipo de análise estática que aplicam regras
de forma mais superficial do que os verificadores de tipo. São regras
relacionadas a espaços em branco, nomes, funções depreciadas, comentários,
estrutura do programa, entre outros.  A maioria dos verificadores de estilo
costumam ser bem flexíveis quanto ao conjunto de regras que aplicam uma vez que
os programadores costumam ser bastante apegados aos seus próprios estilos de
programação. Os erros reportados por verificadores de estilo são aqueles que
afetam a leitura e a manutenabilidade do código-fonte, não indicando potenciais
erros em tempo de execução como fariam os verificadores de tipo.

\subsection{Compreensão de programas}

Ferramentas de compreensão de programa ajudam programadores a terem uma visão
clara frente a grandes programas de computador, ou seja, programas com alto
volume de código-fonte. Ambientes de desenvolvimento integrados (IDE)
geralmente incluem funcionalidade de compreensão, por exemplo, ``encontrar
todos os usos de um certo método'' ou ``encontrar a declaração de uma variável
global''. Análises mais avançadas chegam a incluir, por exemplo, refatoração
automática. Estas ferramentas de compreensão também são úteis para
programadores interessados em entender código-fonte escrito por outros
programadores.

\subsection{Verificação de programas}

Ferramentas de verificação de programa aceitam como entrada uma especificação e
um conjunto de código-fonte e tenta provar que o código está deacordo com a
especificação. Quando a especificação é uma descrição completa de todo o
programa, a ferramenta de verificação poderá realizar uma checagem de
equivalência para garantir que o código-fonte e a especificação combinam de
forma exata. Programadores raramente têm acesso a uma especificação detalhada
suficientemente para ser usada numa checagem de equivalência, o trabalho de
criar esta especificação pode ser maior do que o trabalho de escrever o próprio
código-fonte do programa, desta forma este tipo de verificação formal raramente
acontece. Sendo mais comum a verificação em relação a uma especificação parcial
que detalha apenas parte do comportamento do programa. Isto costuma ser chamado
de verificação de propriedade, grande parte das ferramentas de verificação de
propriedade funcionam aplicando inferências lógicas ou verificação de modelos.

\subsection{Localização de bugs}

O propósito de uma ferramenta de localização de bugs não está em questões de
formatação, como é a verificação de estilo, nem em realizar uma exaustiva e
completa comparação contra uma especificação, como uma ferramenta de
verificacao de programa. Ao invés disso, um localizador de bugs está preocupado
em apontar locais onde o programa, possivelmente, irá se comportar de forma
inesperada. A maioria das ferramentas de localização de bugs são fáceis de usar
porque costumam vir com um conjunto de regras (\textit{bug idioms}) para descrição
de padrões de código que indicam bugs.  Algumas destas ferramentas costumam
usar os mesmos algoritmos utilizados por ferramentas de verificação de
propriedade.

\subsection{Avaliação de segurança}

Ferramentas de análise estática para segurança usam as mesmas técnicas
encontradas nas outras ferramentas, mas por ter um propósito diferente,
identificar problemas de segurança, aplicam estas técnicas de forma diferente.
As primeiras ferramentas de segurança (ITS4, RATS, Flawfinder) eram pouco mais
do que um \textit{``grep''} melhorado; na maior parte, elas escaneavam o código
procurando por funções como por exemplo \textit{``strcpy()''} que são facilmente
usadas de forma inadequada e devem ser inspecionadas manualmente no processo de
revisão de código-fonte. A evolução deste tipo de ferramenta de segurança levou
a técnicas híbridas de verificação de propriedade e de localização de bugs, de
forma que muitas propriedades de segurança podem ser suscintamente expressas
como propriedades de programas.

\section{Anatomia da análise de código-fonte} \label{anatomia}

\begin{figure}[h]
  \center
  \includegraphics[scale=0.4]{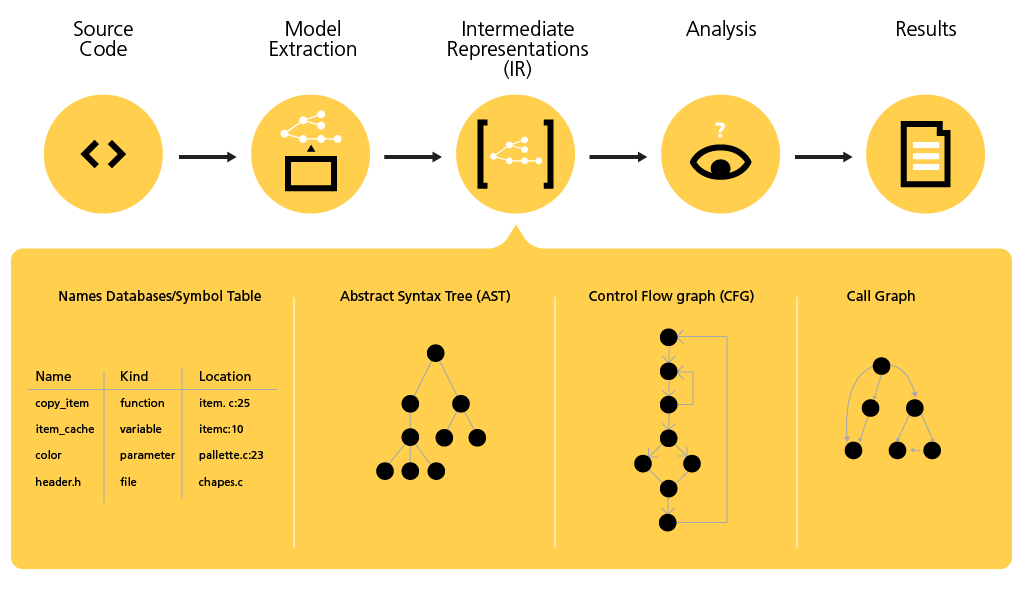}
  \caption{CodeSonar: Representação da Análise Estática \cite{GrammaTech2016}}
  \label{static-analysis-representation}
\end{figure}

Ferramentas de análise estática de código-fonte estão organizadas em partes ou
componentes, responsáveis por implementar três funções básicas \cite{Cruz2009,
Binkley2007}: a) extração de dados, b) geração de representação intermediária,
e c) análise. A ferramenta de análise de código-fonte
CodeSonar\footnote{https://www.grammatech.com/products/codesonar}, por exemplo,
segue tal organização e realiza cada uma das 3 funções em etapas distintas,
conforme mostra a Figura \ref{static-analysis-representation}.

\subsection{Extração de dados}

O processo de recuperar dados para futuro processamento ou armazenamento é
chamado de extração de dados, é o
primeiro componente da análise de código-fonte e é
responsável por ler o código-fonte do programa e gerar uma ou mais
representações intermediárias. Em essência, este componente converte a sintaxe
de um programa em uma outra sintaxe abstrata e mais adequada para análise
posterior. Este componente é usualmente chamado de analisador sintático (ou
\textit{parser}) e apesar de teoricamente não ser uma tarefa difícil, apresenta
enormes desafios diante da complexidade das linguagens de programação modernas.

\subsection{Representação intermediária}

Exportar os dados extraídos para uma representação intermediária é uma
estratégia comum para facilitar análise e transformação de dados e
possivelmente adição de metadados.

Os dados obtidos na extração precisam ser representados em um formato mais
abstrato. Esta é a responsabilidade do segundo componente da análise de
código-fonte: armazenar os dados coletados usando uma representação
intermediária em formato mais adequado para análise automática, abstraindo
aspectos particulares do programa e da linguagem de programação.

Alguns tipos de representação intermediária têm sua origem na área de
compiladores; algumas delas são produzidas diretamente pelo \textit{parser}
enquanto outras requerem uma análise específica. Os formatos mais comuns são
geralmente baseados em grafos. Alguns formatos comumente utilizados são:

\begin{multicols}{2}
  \begin{itemize}
    \item Árvore sintática abstrata
    \item Grafo de fluxo de controle
    \item Árvore sintática abstrata decorada
    \item Grafo de dependência de módulos
    \item Atribuição estática única
    \item Grafo de dependência de valores
  \end{itemize}
\end{multicols}

Estas representações podem ser utilizadas tanto na análise estática quanto na
análise dinâmica. O uso de um ou outro formato depende do tipo de análise e seu
propósito. Pode-se combinar diferentes tipos no sentido de enriquecer e
estruturar a informação extraída.

\subsection{Análise}

Este componente é responsável por realizar inferências a partir dos dados
representados internamente. O processo requer que as informações armazenadas
estejam interconectadas e também interrelacionadas com conhecimento anterior.
Esta análise pode gerar conhecimento quantitativo ou qualitativo, como, por
exemplo, métricas de software ou mineração de dados, respectivamente. Técnicas
de visualização podem ser usadas para apoiar este processo.

Diversas técnicas foram desenvolvidas ao longo do tempo para realizar análise,
algumas delas são brevemente descritas na seção \ref{tecnicas}.

\section{Formatos de representação intermediária} \label{formatos}

Essencialmente, um formato de representação intermediária é uma abstração
precisa das propriedades de um programa representado em um domínio menor. Os
compiladores normalmente constroem esta representação a fim de possuir um
modelo do programa sendo compilado, é comum que compiladores utilizem diversos
formatos durante o curso da compilação.

Em ferramentas de análise estática estes formatos são utilizados durante a fase
de análise para cumprir diversos objetivos, como por exemplo, calcular métricas
de código-fonte. A métrica de complexidade ciclomática de McCabe
\cite{McCabe1976}, por exemplo, é definida com base no grafo de fluxo de
controle (\textit{Control Flow Graph - CFG}) do programa com o seguinte cálculo
$CC = e - n + 2p$. Onde: \textbf{e} é o número de arestas; \textbf{n} é o número de
nós; e \textbf{p} é o número de componentes fortemente conectados no grafo.

Assim, percebe-se que cada formato de representação intermediária
\cite{Nielson2015, Stanier2013, Cruz2009, Ramalho1996} pode ter fins e
objetivos bastante distintos.

\subsection{Árvore sintática abstrata}

A árvore sintática abstrata (AST - Abstract Syntax Tree) representa um programa
tratando os elementos do código-fonte como operadores e operandos organizados
em nós numa árvore, este formato de representação é muito popular em tradutores
\textit{source-to-source}\footnote{http://en.wikipedia.org/wiki/Source-to-source\_compiler}.

\subsection{Grafo de fluxo de controle}

O grafo de fluxo de controle (CFG - Control Flow Graph ou Call Graph) é um
grafo direcionado representando a estrutura de controle de um programa e sua
sequência de instruções, onde as arestas mostram os possíveis caminhos de
execução. Este formato é amplamente utilizado em métodos formais para
otimização de código-fonte.

\subsection{Grafo de fluxo de dados}

O grafo de fluxo de dados (DFG - Data Flow Graph) é também um grafo direcionado
onde as arestas representam o fluxo de dados entre as operações do programa,
este formato pode ser visto como um companheiro do grafo de fluxo de controle
(CFG) e pode ser gerado ao longo de uma mesma análise.

\subsection{Árvore sintática abstrata decorada}

Árvore sintática abstrata decorada (DAST - Decorated Abstract Syntax Tree) é
uma árvore sintática abstrata (AST) melhorada através de um processo de
definiçao de atributos para os símbolos do programa de forma declarativa com
uso de uma Gramática de
Atributos\footnote{https://en.wikipedia.org/wiki/Attribute\_grammar}.

\subsection{Grafo de dependência de módulos}

O grafo de dependência de módulos (MDG - Module Dependence Graph) é um grafo
onde os módulos são representados como nós e as arestas representam as relacões
entre eles, indicando dependência entre os mesmos.

\subsection{Atribuição estática única}

Atribuição estática única (SSA - Static Single Assignment) pode ser vista como
uma variação ou uma propriedade de outros formatos de representação
intermediária, é um método que faz cada variável ser atribuída apenas uma única
vez, facilitando a descoberta de informaçoes sobre os dados representados.

\subsection{Grafo de dependência de valores}

O grafo de dependência de valores (VDG - Value Dependence Graph) é uma variação
que melhora (ao menos para algumas análises) os resultados obtidos a partir da
atribuição estática única (SSA). Ele representa tanto o fluxo de controle
quanto o fluxo de dados e assim simplifica a análise.

\cite{Zaytsev2015} enumera e descreve uma quantidade significativa de
grafos de representação interna para programação funcional.

\section{Técnicas de análise} \label{tecnicas}

Inúmeras técnicas e métodos distintos podem ser utilizados pelas ferramentas
de análise estática, seja com o objetivo de verificação de tipos, localização
de bugs, compreensão de programas, avaliação de segurança, ou outra finalidade
qualquer. Segundo \cite{German2003, Li2010, Hofer2010} as técnicas e
métodos mais comumente encontrados nas ferramentas atuais são apresentados a seguir.

\begin{description}

  \item \textit{Análise léxica}.
    A análise léxica é responsável por quebrar o programa em pequenos fragmentos
    (ou \textit{tokens}) e verificar se estes fragmentos são palavras válidas
    para uma dada linguagem. A análise léxica não leva em consideração a
    sintaxe do programa, sua semântica ou a interação entre módulos.

  \item \textit{Combinação de padrões de texto}.
    A combinação de padrões de texto (\textit{Text-based Pattern Matching}) é a
    maneira mais simples e rápida de procurar vulnerabilidades num código-fonte.

  \item \textit{Inferência de tipos}.
    A inferência de tipos (\textit{Type inference}) refere-se a identificar o
    tipo de variáveis e funções e avaliar se o acesso a elas está em
    conformidade com as regras da linguagem. Linguagens de programação com
    sistema de tipagem incluem mecanismos deste tipo de análise.

  \item \textit{Análise de fluxo de dados}.
    A análise de fluxo de dados (\textit{Data flow analysis}) resume-se a coletar
    informação semântica do código-fonte do programa, e com métodos algébricos
    deduzir a definição e uso das variáveis em tempo de compilação. O objetivo
    é mostrar que nenhum caminho de execução do programa acessa uma variável
    sem definição ou atribuição prévia.

  \item \textit{Verificação de regra}.
    A verificação de regra (\textit{Rule checking}) consiste em checar a segurança
    do programa através de um conjunto de regras pré-estabelecidas.

  \item \textit{Análise de restrição}.
    A análise de restrição (\textit{Constraint analysis}) consiste em gerar
    e resolver restrições no processo de análise de um programa.

  \item \textit{Comparação de caminho}.
    Comparação de caminho (\textit{Patch comparison}) inclui comparação de caminho de
    código-fonte e de código-binário e é usada principalmente para encontrar
    brechas de vulnerabilidade já ``conhecidas'' e previamente divulgadas por
    fornecedores e praticantes da indústria de software.

  \item \textit{Execução simbólica}.
    A execução simbólica (\textit{Symbolic execution}) é usada para representar
    as entradas de um programa através do uso de valores simbólicos ao invés
    de dados reais, produz expressões algébricas sobre a entrada dos símbolos
    do programa durante o processo de implementação e pode detectar
    possibilidade de erros.

  \item \textit{Interpretação abstrata}.
    Interpretação abstrata (\textit{Abstract interpretation}) é uma descrição
    formal da análise do programa. Pelo fato de apenas controlar atributos de
    programa de preocupaçao dos usuários, a interpretação da análise semântica
    é similar ao seu significado semântico real.

  \item \textit{Prova de teoremas}.
    Prova de teoremas (\textit{Theorem proving}) é baseada na análise semântica do
    programa. Converte o programa em fórmulas lógicas e então tenta provar que
    o programa é um teorema válido usando regras e axiomas.

  \item \textit{Verificação de modelo}.
    O processo de verificação de modelos (\textit{Model checking}) primeiro constrói
    um modelo formal do programa tal como uma máquina de estados ou um grafo
    direcionado, então examina e compara o modelo para verificar se o sistema
    cumpre as características pré-definidas. Esta técnica requer a definição e
    descrição das propriedades que devem ser verificados por um pedaço de
    software.

  \item \textit{Verificação formal}.
    Verificação formal (\textit{Formal Checking} ou \textit{Compliance Analysis}) é o
    processo de provar de forma automatizada que o código do programa está
    correto em relação a uma especificação formal dos seus requisitos.

  \item \textit{Análise de fluxo da informação}.
    Análise de fluxo da informação (\textit{Information Flow Analysis}) identifica
    como a execução de uma unidade de código cria dependência entre entradas e
    saídas.

  \item \textit{Verificação de sintaxe}.
    Verificação de sintaxe (\textit{Syntax Checks}) tem o objetivo de encontrar
    violação de regras tais como expressões mal-formadas ou fora do padrão.

  \item \textit{Verificação de intervalo}.
    A análise de verificação de intervalo (\textit{Range Checking}) tem o objetivo
    de verificar que os valores dos dados permanecem dentro de intervalos
    especificados, bem como manter a precisão especificada.

\end{description}

Diante a variedade e a constante evolução da área de análise estática
\cite{Novak2010} fez um estudo propondo uma taxonomia e um conjunto de
dimensões para caracterização de ferramentas de análise estática, lá é
possível obter uma visão ampla e abrangente dos conteitos e temas da área.

\section{Ferramentas de análise estática}

A variedade de aplicação e a constante evolução da área de análise estática,
tanto na indústria quando na academia, resulta em  estudos teóricos e práticos,
novas ferramentas, modelos e algoritmos de análise estática. Ferramentas de
análise estática têm sido continuamente desenvolvidas e seu uso se tornado
comum no ciclo de desenvolvimento de software.

Análise estática é a técnica mais amplamente utilizada para análise
automatizada de programas devido a sua eficiência, boa cobertura e automação.
Estudos mostram que analise estática tem grande adoção em projetos de software
livre \cite{beller2016analyzing}.
Entretanto, técnicas de análise estática amplamente adotadas na comunidade de software,
por exemplo, para localização de bugs e verificação de programas 
ainda sofrem um alto índice de falso-positivos \cite{gosain2015static}.

Apesar da ampla adoção de ferramentas de análise estática em estudos acadêmicos
e da crescente atenção que as técnicas de análise estática de código tem
recebido em pesquisas, nota-se ainda uma enorme distância entre a atenção dada
na academia e sua adoção na indústria, identificando um \textit{gap} entre
estes dois contextos \cite{ilyas2016static}.

Mesmo diante das dificuldades, pesquisadores e cientistas continuam
contribuindo com ferramentas de análise estática, o seu reconhecimento, apesar
de tímido, vem crescendo, a Tabela \ref{ferramentas} apresenta um conjunto
dessas ferramentas com informação para download e breve descrição de suas
principais funções.  Estas ferramentas precisam ser divulgadas, avaliadas e
acima de tudo apropriadas pela comunidade científica, evoluídas e tratadas como
um bem comum para um avanço geral da qualidade, maturidade e confiabilidade
dessas ferramentas tão importantes para pesquisas nas diversas áreas da
Engenharia de Software.

\begin{table}[htbp]
\caption{Ferramentas de análise estática \cite{costa2018sustainability}}
\begin{tabular}{|l|l|}
\hline
Nome & Descrição \\ \hline
\href{http://svn.cprover.org/wiki/doku.php?id=2ls for program analysis}{2LS} & Análise de terminação para programas C usando resumo interprocedural \\ \hline
\href{http://accessanalysis.sourceforge.net}{AccessAnalysis} & Cálculo de métricas IGAT e IGAM (plugin Eclipse) \\ \hline
\href{http://vsl.cis.udel.edu/civl/}{CIVL} & Framework para verificação de programas concorrentes \\ \hline
\href{http://codeboost.org}{CodeBoost} & Transformação source-to-source para otimização de programas C++ \\ \hline
\href{http://www.cs.ucsb.edu/~bultan/composite/}{CSL} & Verificação de modelos \\ \hline
\href{http://users.ecs.soton.ac.uk/gp4/cseq/files/cseq-0.5.zip}{CSeq} & Transformação source-to-source para programas C concorrentes \\ \hline
\href{http://people.csail.mit.edu/jnear/derailer}{Derailer} & Localização de falhas de segurança em aplicações web \\ \hline
\href{https://github.com/saltlab/dompletion}{DOMPLETION} & Sugestão de código javascript \\ \hline
\href{https://www.dropbox.com/s/glhg8any43lccgm/EJB.zip}{EJB} & (EJB Interceptor Analyzer) Criação de diagramas de sequência \\ \hline
\href{http://sourceforge.net/p/emunity/code/ci/master/tree/}{e-munity} & Verificação de segurança \\ \hline
\href{http://code.google.com/p/error-prone}{Error Prone} & Localização de bugs em código Java construído com o javac \\ \hline
\href{http://www.sed.inf.u-szeged.hu/FaultBuster}{FaultBuster} & Refatoração de code smells \\ \hline
\href{https://github.com/jlopezvi/Flowgen}{Flowgen} & Criação automática de grafos UML \\ \hline
\href{http://modelum.es/trac/guizmo/}{GUIZMO} & Inferência de layout \\ \hline
\href{https://github.com/jrfaller/gumtree}{GumTree} & Comparação de mudanças \\ \hline
\href{http://husacct.github.io/HUSACCT}{HUSACCT} & verificação de conformidade arquitetural \\ \hline
\href{http://indus.projects.cis.ksu.edu}{Indus} & Biblioteca de program slicing \\ \hline
\href{http://jastadd.cs.lth.se/web}{JastAdd} & Análise de código-fonte através da descrição de atributos via gramática \\ \hline
\href{http://vazexqi.github.io/JFlow/}{JFlow} & Transformação source-to-source \\ \hline
\href{http://bogor.projects.cs.ksu.edu/manual/}{Bogor/Kiasan} & Verificação de modelos \\ \hline
\href{http://verify.inf.usi.ch/content/loopfrog}{Loopfrog} & Verificação de modelos \\ \hline
\href{https://github.com/MaxLillack/Lotrack}{Lotrack} & Análise estática de configuração \\ \hline
\href{https://github.com/YoshikiHigo/MPAnalyzer}{MPAnalyzer} & Análise de padrões disponível \\ \hline
\href{http://mygcc.free.fr}{mygcc} & Verificação de programas C \\ \hline
\href{https://github.com/cwi-swat/php-analysis}{PHP AiR} & Um framework para análise de código PHP escrito em Rascal \\ \hline
\href{https://github.com/jensnicolay/jipda/tree/scam2015/protopurity}{protopurity} & Análise de impacto \\ \hline
\href{http://ahclab.naist.jp/pseudogen}{Pseudogen} & Transformação de código-fonte em pseudo-código \\ \hline
\href{http://www.cs.toronto.edu/~tomhart/ptyasm}{PtYasm} & Verificação de modelos \\ \hline
\href{http://mir.cs.illinois.edu/reassert}{ReAssert} & Localização de falhas em testes e refatoração \\ \hline
\href{http://github.com/FrontEndART/SonarQube-plug-in}{Sonar Qube Plug-in} & Extensão do SourceMeter para análise de código Java com o SonarQube \\ \hline
\href{http://types.cs.washington.edu/sparta}{SPARTA} & Segurança pra detecção de malware \\ \hline
\href{http://www.srcml.org}{srcML} & Transformação source-to-source \\ \hline
\href{http://presto.cse.ohio-state.edu/tacle}{TACLE} & Análise de tipo, construção e visualizaçao de grafos de chamada \\ \hline
\href{http://tebasaki.jp/src}{TEBA} & Transformação source-to-source \\ \hline
\href{http://wala.sourceforge.net/wiki/index.php/Main\_Page}{WALA} & Análise estática de bytecode Java \\ \hline
\href{http://www.cs.kent.ac.uk/projects/wrangler/Home.html}{Wrangler} & Refatoração de código Erlang \\ \hline
\href{http://perso.unamur.be/~cnagy/scam2017-eng}{SQL-CSD} & A Static Code Smell Detector for SQL Queries Embedded in Java Code \\ \hline
\href{https://github.com/rtse-project/extracting-time-automata}{ETA} & Extract Timed Automata from Java programs \\ \hline
\href{https://github.com/guipadua/JTratch}{Jtratch} & Captura de fluxos de excessão para projetos Java usando Exlipse JDT \\ \hline
\href{https://github.com/ecu-pase-lab/mysql-query-construction-analysis}{MQCA} & Identificação de padrões de querys SQL em códigos PHP \\ \hline
\href{https://github.com/v-m/PropagationAnalysis}{PropagationAnalysis} & Software graphs, mutation testing and propagation estimer tools \\ \hline
\href{https://github.com/Mondego/pyreco}{PyReco} & Type discovery for Python \\ \hline
\href{https://github.com/acieroid/scala-am}{Scala-AM} & Abstract Machine Experiments using Scala \\ \hline
\href{https://github.com/EnSoftCorp/SID}{SID} & A Statically-Informed Dynamic (SID) analysis toolbox \\ \hline
\end{tabular}
\label{ferramentas}
\end{table}

\section{Conclusão}

Análise estática é uma área de estudo da Engenharia de Software com um largo
histórico e em constante evolução, possui uma grande contribuição da academia
na construção de ferramentas de software, mas ainda carece de colaboração entre
seus pares num esforço para reduzir as limitações e inconsistências
ainda existentes \cite{alemerien_experimental_2013}, essas
limitações possuem relevância e devem chamar a atenção de todos aqueles
que trabalham com análise estática, visto que podem levar pesquisadores a obter
resultados inconsistentes \cite{lincke_comparing_2008}, isto inclui pesquisas
sobre métricas, qualidade, visualização, arquitetura, testes, manutenção e
diversas outras áreas que tenham como objeto de estudo o software e o seu
principal artefato, o código-fonte.

Assim, reforço a importância divulgar, avaliar e colaborar com as ferramentas
de análise estática apresentadas na Tabela \ref{ferramentas}, projetos de
software acadêmico, desenvolvidos por cientistas durante suas pesquisas,
disponibilizados livremente, e que ainda possuem baixo ou nenhum
reconhecimento.

\bibliographystyle{plain}
\bibliography{joenio2019analiseestatica}

\end{document}